\newif\ifShowKeys
\ShowKeysfalse

\documentclass[11pt,a4paper]{article} 
\pdfoutput=1
\usepackage[no-natbib-sort]{my-jheppub}
\usepackage{amsmath, amssymb}

\usepackage{mathpazo}
\usepackage{bm}
\usepackage{environ}
\usepackage{mathrsfs}
\usepackage{array,arydshln}

\usepackage{graphicx,epsfig}
\usepackage{epic}
\usepackage{youngtab}
\usepackage{float}
\usepackage{color}
\definecolor{maroon}{rgb}{0.8,0.3,0.}

\usepackage{slashed}
\usepackage[nodayofweek]{date time}

\ifShowKeys \usepackage[notcite]{showkeys} \fi

\usepackage{hyperref}

\usepackage{aurical}
\usepackage[T1]{fontenc}

\usepackage{framed}
\definecolor{shadecolor}{RGB}{255, 230, 204}
\newenvironment{claim}{\begin{shaded}\noindent\itshape\ignorespaces}{\end{shaded}}

\allowdisplaybreaks

\newcommand{\be}{\begin{equation}}
\newcommand{\ee}{\end{equation}}

\newcommand{\mc}{\mathcal }

\newcommand{\bv}{\begin{verbatim}}
\newcommand{\ev}{\end{verbatim}}

\newcommand{\la}{\label}

\newcommand{\schr}{Schr\"odinger }
\newcommand{\eps}{\epsilon}
\newcommand{\wt}{\widetilde}
\newcommand{\E}{\text{E}}

\newcommand{\blue}[1]{\textcolor{blue}{#1}}

\title{On the large $\Omega$-deformations in the  Nekrasov-Shatashvili limit
of $\mc N=2^{*}$ SYM}
\author[a,b]{Matteo Beccaria} 

\abstract{
We study the multi-instanton partition functions of the 
$\Omega$-deformed $\mc N =2^{*}$ $SU(2) $ gauge theory in the Nekrasov-Shatashvili (NS) limit. 
They depend on the deformation parameters $\eps_{1}$,  the scalar field expectation 
value $a$, and the hypermultiplet mass $m$.
At fixed instanton number $k$, they are rational functions of $\eps_{1}, a, m$ and we look for
possible regularities that admit a parametrical description in the number of instantons. In each instanton
sector, the contribution to the deformed Nekrasov prepotential  has poles for {\em large} deformation parameters. To clarify the properties of these singularities
we exploit Bethe/gauge correspondence and examine the special ratios $m/\eps_{1}$ at which the 
associated spectral problem is $n$-gap. At these special points 
we illustrate several structural simplifications occurring in  the partition functions.
After discussing various tools to compute the prepotential, we analyze the non-perturbative
corrections up to $k=24$ instantons and 
present various closed expressions for the coefficients of the singular terms.
Both the regular and singular parts of the prepotential are resummed over all instantons and compared
successfully with the exact prediction from the spectral theory of the Lam\'e equation, showing that the 
pole singularities are an artifact of the instanton expansion. The analysis is fully worked out in the 1-gap case,
but the final pole cancellation is proved  for a generic ratio $m/\eps_{1}$ relating it to the gap width of the 
Lam\'e equation.
\vfill }

\affiliation[a]{Dipartimento di Matematica e Fisica Ennio De Giorgi,\\
Universit\`a del Salento, Via Arnesano, 73100 Lecce, 
Italy} 

\affiliation[b]{INFN, Via Arnesano, 73100 Lecce, Italy}

\emailAdd{matteo.beccaria@le.infn.it} 

\begin{document}

%\date{\currenttime}
%\begin{flushleft}\boxed{\small{\tt \today \ \ - \ \  \currenttime }}\end{flushleft}

% \begin{flushright}\small{Imperial-TP-AT-2015-{06}}\end{flushright}				% report number

\maketitle
\flushbottom
\section{Introduction}

Four-dimensional gauge theories with rigid $\mc N=2$ supersymmetry play a central role
in modern theoretical physics. Roughly speaking, they stand  between the {\em real} world of phenomenological 
applications of softly broken $\mc N=1$ models, and the highbrow realm of maximal $\mc N=4$ theories.
The theoretical control over $\mc N=2$ super Yang-Mills (SYM) theories is largely 
based on the key fact that their low-energy effective physics 
is captured by the Seiberg-Witten (SW) curve. It describes non-perturbatively the geometry of the moduli
space of the gauge theory vacua \cite{Seiberg:1994rs,Seiberg:1994aj}. In the simplest case when the 
gauge group is $SU(2)$, the SW curve is a complex torus whose modular parameter $\tau$ is the 
complexified gauge coupling constant  at low energy. Quite generally, the effective action 
has 1-loop perturbative corrections and instanton non-perturbative corrections. It may be fully
represented in terms of an analytic function, the prepotential $\mc F$. It 
depends on the vacuum expectation value $a=(a_{1}, a_{2}, \dots)$ of the scalars in the adjoint gauge multiplet and on the 
masses $m=(m_{1}, m_{2}, \dots)$ of the hypermultiplets \cite{D'Hoker:1999ft}. 
More structure is provided by combining $\mc N=2$ supersymmetry with conformal invariance.
Here, we shall focus on a standard benchmark, {\em i.e.} the so-called $\mc N=2^{*}$ theory whose matter 
content is an adjoint massive hypermultiplet. The gauge coupling runs at 1-loop by perturbative corrections
proportional to the hypermultiplet masses. The most interesting effects are due to the instantons and are of course
non-perturbative. Our analysis will be aimed at studying 
of these contributions leaving aside the perturbative part.

\medskip
The non-perturbative corrections are predicted correctly by the SW curve, but may be computed in an
alternative and somehow more
powerful way by means of localization \cite{Nekrasov:2002qd,Nekrasov:2003rj}. 
The $\mc N=2$ partition function is topologically twisted and localized on the multi-instanton moduli spaces.
These are finite-dimensional spaces, but nevertheless are plagued by divergences that appear in the UV at small instanton size and 
also in the IR when the instanton centers are moved to infinity.
 These problems 
have been cleverly by-passed by  considering the gauge theory in a curved background, known 
as $\Omega$-background \cite{Nekrasov:2002qd,Nekrasov:2003rj,Nakajima:2003uh}.
This construction introduces two deformation parameters $\eps_{1}, \eps_{2}$ that 
break 4d Poincar\' e invariance but fully regularize the moduli space integration
\cite{Flume:2002az,Bruzzo:2002xf,Flume:2004rp,Nekrasov:2004vw,Marino:2004cn,
Billo:2009di,Fucito:2009rs,Billo:2010bd}.~\footnote{Quite remarkably, 
the $\Omega$-background is equivalent to 
a supergravity background with a non-trivial graviphoton field strength, 
on the instanton moduli space \cite{Billo:2006jm,Ito:2010vx}.
For the string interpretation of the $\Omega$-background and its BPS excitations
see \cite{Hellerman:2011mv,Hellerman:2012zf,Orlando:2013yea}.}
The resulting Nekrasov instanton partition function $Z_{\text{inst}}(\eps_{1},\eps_{2},a,m)$ defines 
a non-perturbative $\epsilon$-deformed prepotential by means of the relation
\be
\la{1.1}
F_{\text{inst}}(\eps_1,\eps_2, a, m)= -\epsilon_1\epsilon_2\,\log Z_{\mathrm{inst}}(\eps_1,\eps_2, a, m).
\ee
The SYM prepotential $\mc F(a,m)$ is recovered in the $\eps_{1},\eps_{2}\to 0$ limit, up to the classical 
tree-level contribution.

\medskip
There are several reasons to regard the $\Omega$-deformation as more than a mere
regularization. {\em i.e.}
to study the properties of (\ref{1.1}) at finite deformation parameters $\eps_{1}, \eps_{2}$. 
One string motivated 
analysis examines  systematically the expansion of
(\ref{1.1}) around the undeformed point $\eps_{1}=\eps_{2}=0$. \footnote{
Indeed, it is convenient to add to (\ref{1.1}) the perturbative 
part $F_{\text{pert}}$ and define the amplitudes $F^{(n,g)}$ from the double expansion
$F_{\text{pert}}+F_{\text{inst}}=\sum_{n,g=0}^\infty F^{(n,g)}\,(\eps_1+\eps_2)^{2n}\,
(\eps_1\,\eps_2)^g$. 
%The amplitude $F^{(0,0)}$ equals the undeformed SYM prepotential $\mc F$,
%up to the classical tree-level term. 
The amplitudes $F^{(0,g)}$ with $g\geq 1$ 
correspond to F-terms in the effective action of the form $\mc W^{2g}$, 
where $\mc W$ is the chiral Weyl superfield containing the graviphoton field strength. 
These terms may also be obtained from the genus $g$ partition function of the $\mc N=2$
topological string  \cite{Antoniadis:1993ze} and satisfy a holomorphic anomaly equation \cite{Bershadsky:1993ta,Bershadsky:1993cx,Klemm:2002pa,Huang:2009md}. Similar interpretations 
can be extended to  the amplitudes $F^{(n,g)}$ with $n\neq 0$ 
\cite{Krefl:2010fm,Huang:2010kf,Antoniadis:2013mna,Antoniadis:2013epe,Florakis:2015ied},
see also \cite{Antoniadis:2010iq}.
}
Another option is to look at the exact dependence of  (\ref{1.1}) on its parameters, including 
$\eps_{1}, \eps_{2}$. Ideally, one would like to emphasize special features
appearing order by order in the instanton expansion. In this perspective, a major simplification 
occurs in the 
Nekrasov-Shatashvili (NS) limit \cite{Nekrasov:2009rc} where one of the two $\epsilon$ parameters
vanishes, say $\eps_{2}=0$. The resulting theory has a 2d $\mc N = 2$ super-Poincar\'e invariance. 
The supersymmetric vacua of this gauge theory are the eigenstates of the quantum integrable system obtained by 
quantization of the classical  integrable system associated with  the geometry of the 
moduli space of undeformed $\mc N = 2$ theory. Under this Bethe/gauge map, the deformation 
parameter $\eps_{1}$ plays the role of the Planck constant. In the NS limit, saddle point methods
permit to derive a generalized SW curve \cite{Poghossian:2010pn,Fucito:2011pn}. \footnote{
A matrix model approach 
to deformed SW theory is developed in \cite{Marshakov:2010fx,Mironov:2009dv,Mironov:2009uv,Bourgine:2012gy,Bourgine:2012bv}.
}

\medskip
Another pivotal tool to control the generalized prepotential (\ref{1.1}) 
non-perturbatively in $\eps_{1},\eps_{2}$
is the AGT correspondence  \cite{Alday:2009aq} that maps 
$\eps$-deformed $\mc N=2$  instanton partition functions  to conformal blocks of a suitable CFT. The details of the gauge
theory (gauge group and matter content) determine the worldsheet genus and 
the number and conformal weights of the insertions in the relevant conformal block. The AGT correspondence
can be checked by comparing the expansion in instanton number with an expansion in a 
complex structure parameter of the corresponding punctured Riemann surface on the CFT side
\cite{Poghossian:2009mk,Fateev:2009aw,Alba:2010qc}. In the specific case of the 
$\mc N=2^{*}$ $\eps$-deformed
theory  with gauge group $SU(2)$, the relevant CFT object is the one-point conformal block on the torus.
An important suggestion from the AGT correspondence is that the conformal blocks have nice modular properties
with a possible counterpart in the gauge theory as a generalized $S$-duality at non zero $\eps_{1}, \eps_{2}$.
\footnote{Notice that the SW contour integral methods 
remain valid also when both $\eps_1$ and $\eps_2$ are non-vanishing \cite{Gaiotto:2009we,Mironov:2009ib}.}
Indeed, modular transformation properties are well known in the undeformed case  \cite{Minahan:1997if,Billo:2011pr}
and may be explored in the deformed case \cite{Huang:2011qx,Huang:2012kn}.
For conformal gauge theories with a single gauge group, such as $SU(2)$ theory with $N_{f} = 4$ 
and the $\mc N = 2^{*}$ theory, it has been possible to make a lot of progress in 
resumming the instanton expansion by expressing the (deformed) prepotential in terms of 
quasi-modular functions (Eisenstein series) of the instanton coupling constant. 
This has been done both from the gauge theory perspective 
 \cite{Billo:2013fi,Billo:2013jba,Billo:2014bja,Billo:2015ria,Billo:2015jta,Billo:2016zbf,Ashok:2016oyh}
and, exploiting AGT, on the CFT side
 \cite{KashaniPoor:2012wb,Piatek:2013ifa,Kashani-Poor:2013oza,Kashani-Poor:2014mua}.
An alternative route exploits an exact WKB analysis
 \cite{Mironov:2009dv,Mironov:2009uv,Mironov:2009dv,He:2010xa,He:2010if,Popolitov:2013ria,He:2014yka}.
 \footnote{
 The WKB analysis is also interesting because it allows to access non-perturbative aspects of the 
 $\eps_{1}$ expansion, see for instance 
 \cite{Krefl:2013bsa,Krefl:2014nfa,Basar:2015xna, Kashani-Poor:2015pca,Ashok:2016yxz}.}
 
 \medskip
 Despite being a powerful tool, quasi-modular expansions are commonly studied after expanding 
 the prepotential in inverse powers of (the components of) $a$, a natural setup from the 
 point of view of the SW curve. As a general feature, the coefficient of each power of $1/a$ 
 turns out to be a polynomial in Eisenstein series and $\theta$-functions with definite modular properties
 that reflect the $SL(2, \mathbb{Z})$ symmetry of the high energy and extend S-duality to the effective theory.
These coefficients have also a polynomial dependence on $\eps_{1}, \eps_{2}$ and the hypermultiplet masses $m$.
So, in a sense, this large $a$ expansion amounts to an educated 
reshuffling of the small $\eps_{1}, \eps_{2}$ expansion.
Instead, at finite parameters $\eps_{1}, \eps_{2}, a, m$ the Nekrasov partition functions $Z_{k}$
at instanton number $k$
is computable as a rational functions of them. On one hand, this means that there
 is a lot of structure to be revealed and 
understood. On the other hand, it seems hopeless to look for regularities at increasing $k$ due to the 
growing complexity of $Z_{k}$.

\medskip
At least in the NS limit, these difficulties may be overcome by tuning the hypermultiplet mass $m$ and
taking it to be proportional to $\eps_{1}$ with definite special ratios
$\tfrac{m}{\eps_{1}} = n+\tfrac{1}{2}$, where $n\in\mathbb{N}$. 
In the  $\mc N=2^{*}$ theory, these points are the $n$-gap cases of the associated 
spectral problem predicted by the Bethe/gauge correspondence, {\em i.e.} the elliptic Calogero-Moser system 
\cite{Nekrasov:2009rc}. 
At these special points, 
we shall discuss a remarkable simplification in the functions $Z_{k}$. In the simplest 1-gap case, they can be reduced
to  rational functions of a single scaled variable and their denominator is a simple power of the factor 
$4\,a^{2}-\eps_{1}^{2}$. Thus the complexity of the computation
is definitely comparable with the large $a$ approach where
modularity was discovered. Besides, at the technical level, 
 the functions $Z_{k}$ may be obtained by expanding 
the eigenvalues of a Lam\'e equation in terms of its Floquet exponent, the quasi-momentum. \footnote{
This construction has an AGT counterpart as a modular expansion of the conformal blocks computed by 
null vector decoupling equations.}

\medskip
Apart from the general Bethe/gauge framework leading to the calculation of $Z_{k}$,
the interesting goal is to  identify which features of $Z_{k}$ may be described
parametrically in $k$, {\em i.e.} at all instanton numbers. Our investigation will show that a special role is 
played by the pole singularities of $Z_{k}$ in the apparently singular 
{\em large} $\Omega$-deformation $\eps_{1}\to 2a$.
It is not easy to understand these poles from the gauge theory point of view. In the $S^4$ partition function, for example, $a/\epsilon$ is necessarily taken to be imaginary so $4a^2-\epsilon_1^2$ is never zero. However, if we forget about the reality properties of $a$ or $\epsilon$, then indeed the Nekrasov partition function $Z_{k}$ have pole
singularities at the above points. To give an interpretation,
%From the point of view of  the dual 2d CFT, 
%they signal the existence of null states. 
some hints come from the analysis of the 5d origin of these poles in the 
$\Omega$-deformation construction \cite{Nekrasov:2004vw,Tachikawa:2004ur}.
Basically, the instanton partition functions at a certain instanton number $k$ may be computed in a reduced supersymmetric quantum mechanical model with hamiltonian $H_{k}$.
They arise in the 4d $\beta\to 0$ limit of twisted partition functions 
$\text{Tr}(-1)^{F}\,e^{-\beta\,H_{k}}\,e^{\beta\,\Gamma}=\text{Tr}_{\mc H^{(k)}_{{\rm susy}}}(-1)^{F}\,e^{\beta\,\Gamma}$, where $\mc H_{\rm susy}$ is the space of supersymmetric states and 
$\beta$ is the radius in the 5d space $S_{\beta}^{1}\times \mc M^{4}$.
The term $\Gamma$ is a linear combination of $a$ and $\eps_{1}$. The poles in $Z_{k}$ appear when $\Gamma=0$
and signal the infinite dimension of the set of supersymmetric states. Their contribution is 
regularized by the $\Gamma$ term  leading to finite sums of the form 
 $\sum_{n} e^{-\beta\,n\,(c\, \eps_{1}+c'\,a)}$ with $n$ over some subset of the integers.~\footnote{The 
 simplest example is that of 
of supersymmetric quantum mechanics on $\mathbb C^{2} = \{(u,v), u,v\in\mathbb C\}$.
If this system has global symmetry generators $(J_{1}, J_{2})$ being $(1,0)$ on $u$ and $(0,1)$ on $v$, then
the partition function $Z(\beta, \eps_{1}, \eps_{2}) = \text{tr}(-1)^{F}e^{i\,\beta\,\eps_{1}\,J_{1}}
e^{i\,\beta\,\eps_{2}\,J_{2}}$ reduces to a sum over $\mc H_{\rm susy} = \oplus_{m,n\ge 0}
\mathbb C\,z^{m}\,w^{n}$ that gives 
$Z(\beta, \eps_{1}, \eps_{2})  = (1-e^{i\,\beta\,\eps_{1}})^{-1}(1-e^{i\,\beta\,\eps_{2}})^{-1}$. In this case,
the singularity is at small $\eps_{1},\eps_{2}\to 0$. In the gauge theory case, we need to add the scalar field vacuum expectation values $a$ playing formally a role similar to that of $\eps_{1}$, $\eps_{2}$
\cite{Tachikawa:2014dja}.} In our setup, $\Gamma=0$
precisely when $\eps_{1}\pm 2\,a$ vanishes. This  specific
origin of the poles suggests that they may be simple enough to be described at all $k$. So, in summary, 
our admittedly vague claim is that
\begin{claim}
The ``large'' $\eps_{1}$ singularities of the $k$-instanton partition functions are closely related to the structure
of $\mc H_{\rm susy}^{(k)}$ and could feature a 
particularly simple dependence on the instanton number $k$.
\end{claim}
\noindent
In this paper, we shall  explore this idea in more 
quantitative terms. To this aim, in Sec.~(\ref{sec:struc}), we shall begin by 
reviewing some special features of the Nekrasov partition functions in the $n$-gap NS limit by analyzing 
explicit examples at low instanton number.
In Sec.~(\ref{3}), we shall review the Bethe/gauge connection for the $\mc N=2^{*}$ theory. In particular, 
we shall illustrate the relation between the Nekrasov functions and the exact properties of the 
Lam\'e equation, as well as various algorithms for the actual computation of $Z_{k}$ for large $k$.
This relatively simple technology will be exploited in Sec.~(\ref{sec:poles-1})
where we shall analyze the explicit 
data for $Z_{k}$ at the 1-gap point considering up to $k=24$ instantons. The main 
outcome will be a set of closed 
expressions providing  the parametric dependence on $k$ of the poles of $Z_{k}$ as $\eps_{1}\to 2a$. 
This will be the quantitative version of the above claim. Besides, 
these singular contributions  admit a resummation at all instantons which turns out to be {\em regular}
at the naively singular point. This finiteness property has a very simple interpretation in terms of the spectral
properties of the Lam\'e equation. The same study is then extended to the 2-gap case
and generalized in Sec.~(\ref{sec:non-gap}) to the generic $\mu$ case.
Remarkably, the finite resummation of the Nekrasov poles turns out to be closely related to the 
Lam\'e equation gap widths.
 
\section{Structure of Nekrasov instanton partition functions}
\la{sec:struc}

The grand-canonical instanton partition function in the $\Omega$-deformed $\mc N=2^{*}$ $SU(2)$ gauge theory may be written as an expansion in the number of instantons according to 
\be
\la{2.1}
Z_{\text{inst}}(\eps_{1}, \eps_{2}, a, m) = 1+\sum_{k=1}^{\infty}Z_{k}(\eps_{1}, \eps_{2}, a, m)\,q^{2k},
\ee
where $q=e^{i\,\pi\,\tau}$ in terms of the complexified gauge coupling $\tau$.
The associated deformed prepotential is, see (\ref{1.1}),
\be
\la{2.2}
F_{\text{inst}}(\eps_{1}, \eps_{2}, 
a, m) = -\eps_{1}\, \eps_{2}\, \log Z_{\text{inst}} = \sum_{k=1}^{\infty} F_{k}(\eps_{1}, \eps_{2}, 
a, m)\,q^{2k},
\ee
where $a$ is the scalar field vacuum expectation value and $m$ is the hypermultiplet mass, as discussed in the Introduction.
The explicit function $Z_{k}$ may be computed by Nekrasov formula \cite{Nekrasov:2003rj} encoding 
localization. In general, for a classical gauge algebra
$\mathfrak g\in \{\text{A}_{r}, \text{B}_{r}, \text{C}_{r}, \text{D}_{r}\}$ one can efficiently apply the 
methods in 
\cite{Nekrasov:2002qd,Nekrasov:2003rj,Bruzzo:2002xf,Fucito:2004ry,Shadchin:2004yx,Marino:2004cn,Billo:2012st}
 to compute (\ref{2.2}), order by order in the instanton number $k$. This is straightforward, but 
  computationally quite involved as $k$ increases, limiting the technique to the first few values of $k$. 
  For our purposes,
it is instructive to look at the explicit contributions $F_{k}$. At one instanton level, 
we have the well-known result
\be
\la{2.3}
F_{1}(\eps_{1}, \eps_{2}, a, m) =-\frac{(4m^{2}-(\epsilon_{1}-\epsilon_{2})^{2})(16a^{2}-4m^{2}-3(\epsilon_{1}+\epsilon_{2})^{2})}{8\,(4a^{2}-(\epsilon_{1}+\epsilon_{2})^{2})}.
\ee
At two instanton level, the expression for $Z_{2}(\eps_{1}, \eps_{2}, a, m)$ is again a rational function of 
its parameters, but much more involved. Apparently, it is hopeless to look for  results parametric in the instanton number $k$. Going to the Nekrasov-Shatashvili limit $\eps_{2}\to 0$ improves little. Just to give a feeling, at $k=2$ we obtain 
the ugly combination
\be
\la{2.4}
\begin{split}
F_{2}&(\eps_{1}, 0, a, m) = \frac{4m^{2}-\eps_{1}^{2}}{1024\,(4a^{2}-\eps_{1}^{2})^{3}\,(a^{2}-\eps_{1}^{2})}
\,\bigg[-256 \,a^2\, \left(192 a^6-96 a^4 m^2+48 a^2 m^4-5 m^6\right)\\
&+64 \,\eps_{1}^{2}\, \left(1248 a^6-288 a^4 m^2+81 a^2 m^4+7 m^6\right)-48\, \epsilon _1^4
   \left(848 a^4-101 a^2 m^2+23 m^4\right)\\
   &+4\, \epsilon _1^6\, \left(2107 a^2-75
   m^2\right)-631 \, \epsilon _1^8\bigg],
   \end{split}
\ee
and again it seems very complicated to generalize to higher $k$. Obviously, the main complications is the increasing complexity of the involved rational functions. In particular, denominators of $F_{k}$ become more and 
more complicated as $k$ grows,
although always in factorized form as in the above examples. 
In the following, to simplify the discussion, we shall name {\bf\em Nekrasov functions} the 
contributions $F_{k}$. Also, we shall name {\bf\em Nekrasov poles} the singularities associated with the vanishing factors in the denominators of the Nekrasov functions.

Looking at (\ref{2.3}) and (\ref{2.4}), it is clear that important simplifications 
occur if we assume that $m\sim \eps_{1}$. Motivated by the following discussion, we shall
parametrize this limit as 
\be
\la{2.5}
m^{2} = \left(\mu+\frac{1}{4}\right)\, \eps_{1}^{2}.
\ee
Dimensional analysis suggests to introduce the quantity
\be
\wt F_{k}(\nu) = \frac{1}{\eps_{1}^{2}}\,F_{k}\bigg(\frac{2a}{\nu}, 0, a, \frac{2a}{\nu}\,\sqrt{\mu+\frac{1}{4}}\bigg).
\ee
This is a function of the scaling variable $\nu$ with an omitted understood 
dependence on the parameter $\mu$. Again, let us look at the first two instanton numbers. We find 
\be
\la{2.7}
\begin{split}
\wt F_{1}(\nu) &= \frac{2 \mu  \left(\mu -\nu ^2+1\right)}{\nu ^2-1}, \\
\wt F_{2}(\nu) &= \frac{\mu  \left(\mu ^3 \left(5 \nu ^2+7\right)-12 \mu ^2 \left(\nu ^2-1\right)^2+6
   \mu  \left(\nu ^2-2\right) \left(\nu ^2-1\right)^2-3 \left(\nu ^2-4\right)
   \left(\nu ^2-1\right)^3\right)}{\left(\nu ^2-4\right) \left(\nu ^2-1\right)^3}.
 \end{split}
\ee
A special case appears immediately, {\em i.e.} $\mu=2$. We shall call this point the {\em 1-gap NS} limit
for reasons that will appear clearly in our later discussion. Computing also $\wt F_{k}$ for $k=3,4,5$ 
at $\mu=2$, we find the following very simple structure of the Nekrasov functions ($Q$ is polynomial)
\be
\la{2.8}
\wt F_{k}(\nu) = \frac{Q_k(\nu^{2})}{(\nu^{2}-1)^{2k-1}}, \qquad \text{deg}\, Q_{k} = 2k-1,
\ee
with the following explicit first cases 
\begin{align}
\la{2.9}
Q_{1}(x) &= -4 (x-3), \qquad
Q_{2}(x) = -2 \left(3 x^3-21 x^2+57 x-7\right),  \\
Q_{3}(x) &= -\frac{16}{3} \left(x^5-11 x^4+94 x^3-226 x^2-111 x-3\right), \notag \\ 
Q_{4}(x) &= -7 x^7+105 x^6-1395 x^5+8909 x^4-10101 x^3-32133 x^2-6353 x+15, \notag \\ 
Q_{5}(x) &= -\frac{8}{5} \, (3 x^9-57 x^8+1668 x^7-21532 x^6+85458 x^5+89754 x^4\notag \\
& -623356
   x^3-413244 x^2-36189 x-9).\notag 
\end{align}
 In other words, in the denominators of $\wt F_{k}$, all factors but one  do cancel and 
 there is only one Nekrasov pole at $\nu^{2}=1$. Thus, 
the analysis of the instanton contribution reduces to the study of the polynomials in (\ref{2.9}).
Actually, there is another important feature that can be checked from the explicit expressions (\ref{2.9})
and that we found to be quite general.
This is a simple {\em selection rule} forbidding even poles around $\nu=1$. In other words, for $\nu\to 1$ \be
\la{2.10}
\wt F_{k}(\nu) = \frac{d_{1}^{(k)}}{(\nu-1)^{2k-1}}+\frac{\blue{0}}{(\nu-1)^{2k-2}}+
 \frac{d_{2}^{(k)}}{(\nu-1)^{2k-3}}+\frac{\blue{0}}{(\nu-1)^{2k-4}}+\cdots+\frac{d_{k}^{(k)}}{\nu-1}+\text{regular},
\ee
and $\wt F_{k}$ is fully determined by the $k+1$ rational numbers 
$\mathbf{d}^{(k)}=\{d_{1}^{(k)}, \dots, d_{k+1}^{(k)}\}$ appearing in the 
{\em exact} decomposition (it includes the regular part)
\be
\la{2.11}\boxed{
\wt F_{k}(\nu) = d_{k+1}^{(k)}+\sum_{p=1}^{k}d_{p}^{(k)}\,\bigg(\frac{1}{(\nu-1)^{2k-2p+1}}-\frac{1}{(\nu+1)^{2k-2p+1}}
\bigg).}
\ee
Our aim will be that of analyzing the coefficients $d_{p}^{(k)}$ in (\ref{2.11}) by considering 
explicit data at high $k$. This can be achieved by exploiting the Bethe/gauge correspondence as a device
to boost the computation of the Nekrasov functions.

\section{Determination of the 1-gap Nekrasov functions from the Lam\'e equation}
\la{3}

As we mentioned, 
the determination of $\wt F_{k}$ from a direct application of Nekrasov formula is not feasible for large $k$
and we heavily exploit Bethe/gauge correspondence. 
The identification  (\ref{2.5})
%with $\mu=n(n+1)$ and $n\in \mathbb N$ 
is precisely associated with the 
integrable model description of the NS limit $\eps_{2}=0$ in terms of the 
elliptic Calogero-Moser system. This further reduces to the  Lam\'e equation in the $SU(2)$ case considered 
here \cite{Nekrasov:2009rc}. Also, when $\mu=n(n+1)$ and $n\in \mathbb N$, the Lam\'e potential is 
finite-gap.
The associated spectral problem is 
\be
\la{3.1}
\psi''(x)-u(x)\,\psi(x)= \lambda\, \psi(x),
\ee
with 
\be
\la{3.2}
u(x) = \mu\,\wp(x; \omega, \omega').
\ee
The Ansatz 
$\psi(x) = \exp(\int^{x} v(x')dx')$ leads to the  Miura equation
\be
\la{3.3}
v'+v^{2} = u+\lambda.
\ee
The Floquet exponent $i\,\nu$ is defined as the average of $v(x)$ over a 
period. \footnote{We shall consider elliptic 
functions $u$ doubly periodic in the complex plane. The average will be taken along the real axis with a possible 
imaginary shift. As long as the integration segment does not cross any pole, the result is independent on this shift.}
The Bethe/gauge correspondence predicts that the full
Nekrasov function $F_{\text{inst}}$ is essentially the energy eigenvalue of the auxiliary \schr equation (\ref{3.1})
expressed in terms of the Floquet exponent $\lambda=\lambda(\nu)$. This relation is in 1-1 correspondence
with the AGT evaluation of the conformal block as an expansion in terms of the inverse Shapovalov
matrix \cite{Mironov:2009dv,Mironov:2009uv,Gaiotto:2009ma}.
Thus, if we evaluate this relation and expand in the modular
parameter $q$ we can obtain immediately all the Nekrasov functions at finite $a$. The 1-gap point $\mu=2$ is 
particularly simple because of the simplicity of the exact Floquet exponent. To work out the details, let us begin by 
denoting
\be
\la{3.4}
\mathbb K = \mathbb K(m), \qquad \mathbb K' = \mathbb K(1-m), \qquad q = \exp\left(-\pi\frac{\mathbb K'}{\mathbb K}\right).
\ee
Then, for $\mu=2$, the potential (\ref{3.2})
can be written  in the Lam\'e form by shifting $x\to x+i\,\mathbb K'$
\be
\la{3.5}
u(x) = 2\,\bigg[-\frac{1+m}{3}+m\,\text{sn}^{2}(x, m)\bigg],
\ee
To map the period interval $[0,4\,\mathbb K]$ to $[0,2\,\pi]$, it is convenient to introduce the scaled
quantities
\be
\nu = \overline\nu\,\frac{2\mathbb K}{\pi}, \qquad \overline\lambda = \lambda\,\bigg(\frac{\pi}{2\,\mathbb K}\bigg)^{2}.
\ee
The {\em exact} Floquet exponent can be written in terms of the Jacobi $Z$ function by the expression
\cite{whittaker1927modern} 
\be
i\,\overline\nu = -i\,\bigg[
i\,Z\left(\arcsin\sqrt{\frac{\overline\lambda+\frac{m+1}{3}}{m}}, m\right)+\frac{\pi}{2\,\mathbb K}
\bigg].
\ee
For our purposes, it will be important to exploit the following representation of $Z$
\be
Z(z, m) = \int_{0}^{\sin z}\frac{1}{\sqrt{1-t^{2}}}\,\bigg(
\sqrt{1-m\,t^{2}}-
\frac{\mathbb E}{\mathbb K\,\sqrt{1-m\,t^{2}}}\bigg)\, dt.
\ee
Separating out real and imaginary parts, we obtain the very useful integral representation of the Floquet exponent
\be
\la{3.9}
i\,\overline\nu = \int_{1}^{\sqrt{\overline\lambda+\frac{m+1}{3}}}\frac{dt}{\sqrt{t^{2}-m}}\,\bigg[
\sqrt{t^{2}-1}+\frac{\mathbb E}{\mathbb K\,\sqrt{t^{2}-1}}
\bigg].
\ee
The integrand can be expanded in series of $m$ and integrated in terms of elementary functions. The resulting expansions -- written for the unbarred quantities $\nu$ and $\lambda$ -- turns
out to be 
\be
\la{3.10}
\begin{split}
i\, \nu &= \sqrt{\lambda-\tfrac{2}{3}}\,\bigg[
1+\frac{3\, (3 \,\lambda -5)}{32\, (3 \,\lambda -2) (3 \,\lambda +1)}\,m^{2}+\frac{3 (3 \,\lambda
   -5)}{32\, (3 \,\lambda -2) (3 \,\lambda +1)}\,m^{3}\\
   &+\frac{3 \left(19035 \,\lambda ^4-32535
   \,\lambda ^3-4725 \,\lambda ^2+5979 \,\lambda +3670\right) }{8192 (3 \,\lambda -2)^2
   (3 \,\lambda +1)^3}\,m^4\\
   &+\frac{3 \left(8667 \,\lambda ^4-15255 \,\lambda ^3-1269 \,\lambda
   ^2+987 \,\lambda +2390\right) }{4096 (3 \,\lambda -2)^2 (3 \,\lambda +1)^3}\,m^5\\
   &+\frac{3}{262144 (3 \,\lambda -2)^3 (3 \,\lambda +1)^5}\,\bigg(
13701555 \,\lambda ^7-24815160 \,\lambda ^6-5143095 \,\lambda ^5\\
&+5885136 \,\lambda
   ^4+7293375 \,\lambda ^3-68742 \,\lambda ^2-1192083 \,\lambda -453050\bigg)\,
   m^6+\dots.
\bigg].
\end{split}
\ee
Notice that this series can be extended with minor effort.
Inverting the expansion (\ref{3.10}), we get a series for the eigenvalue $\lambda$ as a function
of the Floquet exponent
\begin{align}
\la{3.11}
\lambda &= \frac{2}{3}-\nu ^2+\frac{\nu ^2+1}{16(1-\nu
   ^2)}\,m^{2}
   +\frac{\nu ^2+1}{16(1- \nu ^2)}\,m^{3}+\frac{-235 \,\nu ^6+229
   \,\nu ^4+271 \,\nu ^2-233}{4096 \left(\nu ^2-1\right)^3}\, m^4 \notag \\
   & +\frac{-107
   \nu ^6+101 \,\nu ^4+143 \,\nu ^2-105}{2048 \left(\nu
   ^2-1\right)^3}\, m^5\\
   &-\frac{6265 \,\nu ^{10}-18253 \,\nu ^8+8172 \,\nu ^6+19472 \,\nu
   ^4-21613 \,\nu ^2+6085}{131072 \left(\nu
   ^2-1\right)^5}\, m^6+\dots .\notag
\end{align}
Finally, we replace $m$ by $q$ according to (\ref{3.4}), and obtain 
\be
\begin{split}
\la{3.12}
\lambda &= \frac{2}{3}-\nu ^2-\frac{16 \left(\nu ^2+1\right) }{\nu
   ^2-1}\,q^{2}-\frac{16 \left(3 \nu ^6+3 \nu ^4-39 \nu ^2+1\right) }{\left(\nu
   ^2-1\right)^3}\,q^{4}\\
   & -\frac{64 \left(\nu ^{10}+\nu ^8-74 \nu ^6+206 \nu ^4+121 \nu
   ^2+1\right) }{\left(\nu ^2-1\right)^5}\,q^{6}+\dots.
\end{split}
\ee
If we now write (\ref{3.12}) by separating out the $q^{0}$ term  
\be
\la{3.13}
\lambda = \frac{2}{3}-\nu^{2}+\Lambda(\nu, q),
\ee
we can check that the Bethe/gauge correspondence works perfectly in the following form 
\be
\la{3.14}\boxed{
\sum_{k=1}^{\infty}\wt F_{k}(\nu)\,q^{2k} = -\frac{1}{2}\,\int \frac{dq}{q}\,\Lambda(\nu, q)
+8\,\sum_{k=1}^{\infty}\log(1-q^{2k}).}
\ee
In other words, the expansion (\ref{3.12}) in rational functions is  the generating function 
of the Nekrasov partition functions $\wt F_{k}(\nu)$, up to trivial operations. 
The last term in (\ref{3.14}) is related to a $\nu$-independent term appearing in the 
prepotential and proportional to $\log\eta(q)$ where $\eta(q)$ is the Dedekind function, see 
App.~(\ref{app:eisen}). The compact relation  (\ref{3.14}) emphasizes the direct relation that links
$\Lambda$ to the infinite set of Nekrasov functions. Just for the purpose of illustration, 
we can examine (\ref{3.14}) at the two instanton level. It reads
\be
\la{3.15}
\begin{split}
& \wt F_{1}(\nu)\,q^{2}+\wt F_{2}(\nu)\,q^{4}+\dots \\
&=  \frac{4 \left(\nu ^2+1\right) }{\nu
   ^2-1}\,q^{2}+\frac{2 \left(3 \,\nu ^6+3 \,\nu ^4-39 \,\nu ^2+1\right) }{\left(\nu
   ^2-1\right)^3}\,q^{4}+\dots
 +8\,\bigg(-q^{2}-\frac{3}{2}\,q^{4}+\dots\bigg) \\
 &=-\frac{4\,(\nu^{2}-3)}{\nu^{2}-1}\,q^{2}-\frac{2\,(3\,\nu^{6}-21\,\nu^{4}+57\,\nu^{2}
 -7}{(\,\nu^{2}-1)^{3}}\,q^{4}+\dots,
\end{split}
\ee
in full agreement with the expressions of $Q_{1}$ and $Q_{2}$ in (\ref{2.9}).

\section{Algorithms for the Nekrasov functions at generic $\mu$}

The analysis of the 1-gap point $\mu=2$ is important because the Floquet exponent 
can be given in closed form according to (\ref{3.9}). The modular expansion in powers of $q$ is 
just useful to extract the Nekrasov functions order by order in the instanton expansion. 
Here, we present two algorithms to determine the Nekrasov functions at generic $\mu$, {\em i.e.}
at generic ratio $m/\eps_{1}$ between the hypermultiplet mass and the deformation parameter.

\subsection{Modular expansion from continued fraction expansions}

As very nicely observed in \cite{Basar:2015xna}  for the Mathieu equation (see also the recent 
developments \cite{Kashani-Poor:2015pca,Ashok:2016yxz}),
the expansion $\lambda(\nu)$ in (\ref{3.12}) may be
%derived noting that the poles at integer $\nu$ are related to the the fact that $\lambda$ crosses
%the spectral band edges of the Lam\'e equation. Actually, $\lambda(\nu)$ may be 
extracted 
from a known continued fraction expansion of the $N$-th Lam\'e eigenvalue in terms
of $N$ \cite{volkmer2004four}. This  can be used to derive efficiently the expansion (\ref{3.12}) generating the Nekrasov partition function. Notice that this approach works for generic $\mu$, and not only for 
$\mu=n(n+1)$, {\em i.e.} at the $n$-gap points. In algorithmic form, the method is as follows.
We begin by defining $H=2\,L-n(n+1)\,m$, where $n$ is a real parameter not necessarily integer.
We also define the coefficients
\be
\la{4.1}
\begin{split}
\alpha_{p} &= \frac{1}{2}(n-2p-1)(n+2p+2)\,m,\\ 
\beta_{p} &= 4\,p^{2}\,(2-m), \\ 
\gamma_{p} &= \frac{1}{2}(n+2p+2)(n+2p-1)\,m.
\end{split}
\ee
Then, we consider the continued fraction 
\be
\la{4.2}
\beta_{p}-H-\frac{\alpha_{p-1}\,\gamma_{p}}{\beta_{p-1}-H-}\frac{\alpha_{p-2}\,\gamma_{p-1}}{\beta_{p-2}-H-}
\cdots = \frac{\alpha_{p}\,\gamma_{p+1}}{\beta_{p+1}-H-}\frac{\alpha_{p+1}\,\gamma_{p+2}}{\beta_{p+2}-H-}
\cdots\,.
\ee
Expanding $L$ in powers of $m$ with $p$ being a formal parameter, and trading $n$ by $\mu=n(n+1)$,
we obtain 
\begin{align}
\la{4.3}
 L&=
4 p^2+m \left(\frac{\mu }{2}-2 p^2\right)+\frac{m^2 \left((\mu -2) \mu -48
   p^4+4 (2 \mu +3) p^2\right)}{32 \left(4 p^2-1\right)}\\
   &+\frac{m^3 \left((\mu
   -2) \mu -48 p^4+4 (2 \mu +3) p^2\right)}{64 \left(4
   p^2-1\right)}+\frac{m^4}{32768 \left(p^2-1\right) \left(4
   p^2-1\right)^3}\,\bigg[ \notag \\
   &\mu  \left(7 \mu ^3-12 \mu ^2-332 \mu
   +656\right)-248832 p^{10}+256 (164 \mu +1701) p^8\notag \\
   &+64 \left(86 \mu ^2-1148
   \mu -3645\right) p^6-48 \left(4 \mu ^3+168 \mu ^2-820 \mu -1053\right)
   p^4\notag \\
   &+4 \left(5 \mu ^4+24 \mu ^3+750 \mu ^2-2132 \mu -972\right)
   p^2\bigg]+O\left(m^5\right).\notag
\end{align}
Of course it is necessary to keep only a finite number of terms in (\ref{4.2})
for any desired order of expansion.
If we now replace $p\to \nu/2$, we can see that (\ref{3.12}) is reproduced by setting 
\be
\lambda = \bigg(\frac{2\,\mathbb K}{\pi}\bigg)^{2}\,\bigg(-L+\mu\,\frac{m+1}{3}\bigg),
\ee
where the prefactor is the usual one converting the period from $4\mathbb K$ to $2\pi$, and the 
shift in the round bracket is that appearing in  (\ref{3.5}), for generic $\mu$. We now 
express $m$ in terms of $q$ and separate out the 
$q$-independent part according to the generalization of (\ref{3.13})
\be
\la{4.5}
\lambda = \frac{\mu}{3}-\nu^{2}+\Lambda,
\ee
we can check that (\ref{3.14}) reads in general form 
\be
\la{4.6}
\sum_{k=1}^{\infty}\wt F_{k}(\nu)\,q^{2k} = -\frac{1}{2}\,\int \frac{dq}{q}\,\Lambda(\nu, q)
+4\,\mu\,\sum_{k=1}^{\infty}\log(1-q^{2k}).
\ee
For instance, the terms written in (\ref{4.3}) are enough to reproduce perfectly the results (\ref{2.7}).
We conclude by remarking that the expansion of $\lambda(\nu)$
may also be derived by elementary means, although with some
effort as briefly
discussed in App.~(\ref{A}).

\subsection{Matching poles against quasi-modular expansion}

Another method to compute $\wt F_{k}(\nu)$ is based on the general form (\ref{2.8})
taking into account the structure (\ref{2.10}). This is a bit involved at generic $\mu$ and we
illustrate here for the simpler case of $\mu=2$. Nevertheless, the main core of the computation will
be valid for all $\mu$. In the 1-gap case,  at each instanton level, we need the $k+1$ numbers 
$\mathbf{d}^{(k)} = \{d_{1}^{(k)}, \dots, d_{k+1}^{(k)}\}$, see (\ref{2.11}). These may be fixed 
by exploiting the quasi-modular expansion of the prepotential at large $a$.
\footnote{
This may appear to be a backstep from our strategy of avoiding expansions in $a$, but we shall see later 
why we can use this expansion to reconstruct the full rational functions $\wt F_{k}$.}
Let us briefly review the construction, see for instance \cite{He:2014yka}.
We begin by looking for  a solution of the Miura equation (\ref{3.3}) at large $\lambda$. This takes
the form 
\be
\la{4.7}
v = \sqrt\lambda +\sum_{n=1}^{\infty}\frac{v_{n}}{(\sqrt\lambda)^n}.
\ee
Replacing (\ref{4.7}) into (\ref{3.3}), we immediately obtain $v_{n} = v_{n}(u, u', u'', \dots)$. The terms 
with even $n$ are total derivatives. Terms with odd $n$ involve up to the $2(n-1)$-th derivative of $u$.
The first cases are \footnote{The combinations $v_{n}$ are basically KdV charge densities.
From the general properties of the KdV equation \cite{miura1968korteweg}, one has the following recursion
$
v_{n} = -\frac{1}{2}\, v_{n-1}'-\frac{1}{2}\, \sum_{m=1}^{n-2}v_{m}\, v_{n-m-1}
$, with $v_{1} = \frac{u}{2}$.
This is quite direct, but has the disadvantage that one needs the trivial even charges.}
\be
\la{4.8}
\begin{split}
& v_{1} = \tfrac{1}{2}\,u, \ 
v_{2} = -\tfrac{1}{4}\, u', \
v_{3} = \tfrac{1}{8}\,u''-\tfrac{1}{8}\,u^{2}, \
v_{4} = \left(-\tfrac{1}{16}\,u''+\tfrac{1}{8}\,u^{2}\right)', \\
& v_{5} = \tfrac{1}{32}\,u^{(4)}-\tfrac{3}{16}\,u\,u''-\tfrac{5}{32}\,u'^{2}+\tfrac{1}{16}\,u^{3}, \ \dots.
\end{split}
\ee
The Floquet exponent $i\,\nu$ is obtained by taking the average of $v$ over a period (we denote
it by $\langle\cdots\rangle$) giving 
\be
\la{4.9}
i\, \nu = \langle v \rangle = \sqrt\lambda+\sum_{n=1}^{\infty}\frac{\pi^{2n}\, \varepsilon_{n}}{(\sqrt\lambda)^{2n-1}}.
\ee  
Now, we use some specific properties of the 
the potential (\ref{3.2}) suitable for the $\mc N=2^{*}$ theory with gauge group $SU(2)$.
We rescale the real semi-period of the Weierstrass function $\omega\to \omega=1/2$.
The Weierstrass invariants can be expressed in terms of the Eisenstein series $\E_{4}$ and $\E_{6}$,
see App.~(\ref{app:eisen})
\be
\la{4.10}
g_{2} = \frac{4\, \pi^{4}}{3}\, \text{E}_{4}(q), \qquad
g_{3} = \frac{8\, \pi^{6}}{27}\, \text{E}_{6}(q).
\ee
The differential equation for $\wp$ can be used to reduce $v_{n}$ to combinations of powers of $\wp$ with coefficients
involving $g_{2}, g_{3}$.
Defining $p_{n} = \langle \wp^{n}\rangle$, we have 
\be
\la{4.11}
p_{0} = 1, \qquad p_{1} = -\frac{\pi^{2}}{3}\, \E_{2}(q), 
\ee
and the known recursion (see App. A.2 of \cite{KashaniPoor:2012wb} and also \cite{grosset2006elliptic}) \footnote{
The proof of (\ref{4.12}) if  elementary. From the differential equation $(\wp')^{2} = 4\wp^{3}-g_{2}\wp-g_{3}$, we obtain 
$\wp'' = 6\,\wp^{2}-\frac{1}{2}\,g_{2}$, 
and $(\wp^{n-2}\wp')' = (n-2)\, \wp^{n-3}\, (\wp')^{2}+\wp^{n-2}\, \wp'' 
= 2(2n-1)\,\wp^{n}-\frac{1}{2}\,g_{2}\,(2n-3)\,\wp^{n-2}-g_{3}\,(n-2)\,\wp^{n-3}$.
Integrating over a period, we obtain (\ref{4.12}).
}
\be
\la{4.12}
p_{n} = \frac{2n-3}{4(2n-1)}\, g_{2}\, p_{n-2}+\frac{n-2}{2(2n-1)}\, g_{3}\, p_{n-3}. 
\ee
In this way, we obtain the quantities $\varepsilon_{n}$ in (\ref{4.9}). The first cases are \footnote{
As remarked in \cite{Basar:2015xna}, 
the combinations $\varepsilon_{n}$ are essentially Faulhaber polynomials \cite{grosset2006elliptic}.}
\be
\la{4.13}
\begin{split}
\varepsilon_{1} &= -\frac{\mu}{6}\, \E_{2}, \qquad
\varepsilon_{2} = -\frac{\mu^{2}}{72}\, \E_{4}, \\
\varepsilon_{3} &= -\frac{\mu^{3}}{2160}\, (9\,\E_{2}\,\E_{4}-4\,\E_{6})
+\frac{\mu^{2}}{180}\, (\E_{2}\, \E_{4}-\E_{6}), \\
\varepsilon_{4} &= -\frac{5\,\mu^{4}}{72576}\,(15\,\E_{4}^{2}-8\,\E_{2}\,\E_{6})+
\frac{5\,\mu^{3}}{1512}\,(\E_{4}^{2}-\E_{2}\,\E_{6})-\frac{\mu^{2}}{252}(\E_{4}^{2}-\E_{2}\,\E_{6}).
\end{split}
\ee
The expansion (\ref{4.9}) can be inverted in the form 
\be
\la{4.14}
\lambda = -\nu^{2}+\sum_{n=1}^{\infty} \frac{\pi^{2n}\,\lambda_{n}}{\nu^{2(n-1)}},
\ee
with the explicit coefficients
\be
\la{4.15}
\lambda_{1} = \frac{\mu}{3}\,\E_{2}, \quad
\lambda_{2} = \frac{\mu^{2}}{36}\,(\E_{2}^{2}-\E_{4}), \quad
\lambda_{3} = \frac{\mu^{3}}{540}\,(5\E_{2}^{3}-3\E_{2}\E_{4}-2\E_{6})
-\frac{\mu^{2}}{90}(\E_{2}\E_{4}-\E_{6}),
\ee
and so on. We have computed the coefficients $\lambda_{n}$ up to high $n\sim 50$ for later application.
In the NS limit with the identification (\ref{2.5}), the resummed instanton contributions to the 
prepotential can be expanded in inverse powers of $a$ \footnote{
Of course, the full
prepotential has also a perturbative part starting with a RG term $\sim \log(a/\Lambda)$ plus inverse powers of $a$.
Here, we shall focus on the instantons only, {\em i.e} the terms in (\ref{2.2}).} 
and it takes the form \footnote{Many coefficients $f_{n}$ can be 
found in \cite{Billo:2013fi} for generic $m, \eps_{1}$. }
\be
\la{4.16}
F_{\text{inst}}(\eps_{1}, a) = \sum_{k=1}^{\infty} F_{k}(\eps_{1}, a)
\,q^{2k} = \eps_{1}^{2}\,f_{0}(q)
+\sum_{n=1}^{\infty} f_{n}(q)\,\frac{\eps_{1}^{2(n+1)}}{a^{2n}}.
\ee
The relation between $\{f_{n}\}$
and $\{\lambda_{n}\}$ is nothing but Matone relation \cite{Matone:1995rx} in the present Bethe/gauge context. For $n\ge 1$, it simply
reads
\be
\la{4.17}
q\,\frac{d}{dq}\,f_{n}(q)= -\frac{1}{2^{2n+1}}\, \lambda_{n+1}(q).
\ee
This equation must be interpreted in an expansion in powers of $q^{2}$. In particular, the operation 
$(q\frac{d}{dq})^{-1}$ acts on the right side by replacing $q^{n}$ by $\frac{1}{n}q^{n}$ and this fixes
implicitly the integration constant in (\ref{4.17}) when deriving $\{f_{n}\}$ from $\{\lambda_{n}\}$. The term $f_{0}$
can also be worked out and it is 
\be
\la{4.18}
f_{0} = 2\,\mu\,\log\prod_{n=1}^{\infty}(1-q^{2n}).
\ee
As is well known, it is possible to integrate  (\ref{4.17}) by using the differential equations obeyed by the modular functions $\E_{2}, \E_{4}$, and $\E_{6}$
\be
q\frac{d}{dq}\E_{2}(q) = \frac{1}{6}\,(\E_{2}^{2}-\E_{4}), \ \  \ \ \
q\frac{d}{dq}\E_{4}(q) = \frac{2}{3}(\E_{2}\E_{4}-\E_{6}), \ \ \ \ \
q\frac{d}{dq}\E_{6}(q) = \E_{2}\E_{6}-\E_{4}^{2}.
\ee
One finds that $f_{n}$ are  quasi-modular polynomials in $\E_{2}$, $\E_{4}$ and $\E_{6}$ of 
total weight  $2n$. The quasi-modular part is the one involving $\E_{2}$ and is governed by the modular anomaly equation expressing S-duality
 \cite{Billo:2013fi,Billo:2013jba,Billo:2014bja,Billo:2015ria,Billo:2015jta,Billo:2016zbf}.

We now specialize to $\mu=2$ and come back to the problem of determining 
at each instanton level the $k+1$ numbers 
$\mathbf{d}^{(k)} = \{d_{1}^{(k)}, \dots, d_{k+1}^{(k)}\}$, see (\ref{2.11}).
These can be fixed by the modular expansion (\ref{4.14}), using  (\ref{4.16}) and (\ref{4.17}). 
In other words, we
expand 
\be
\la{4.20}
\wt F_{k}(\nu) =  f_{0,k}
+\sum_{n=1}^{\infty} 2^{2n}\,f_{n,k}\,\nu^{-2n} = f_{0,k}-\frac{1}{4\,k}\,\sum_{n=1}^{\infty}
\lambda_{n+1,k}\,\nu^{-2n},
\ee
where $f_{0}(q) = \sum_{k=1}^{\infty} f_{0,k}\,q^{2k}$ and $\lambda_{n}(q) = 
\sum_{k=1}^{\infty} \lambda_{n,k}\,q^{2k}$. Comparing (\ref{4.20}) with the Ansatz  (\ref{2.11})
we determine $\mathbf{d}^{(k)}$.  Just to give an example and illustrate this simple procedure, 
let us consider the 3-instanton case. We have 
\be
\wt F_{3}(\nu) = d_1^{(3)}\,\bigg[\frac{1}{(\nu -1)^5}-\frac{1}{(\nu +1)^5}\bigg]
+d_2^{(3)}\,\bigg[\frac{1}{(\nu -1)^3}-\frac{1}{(\nu +1)^3}\bigg]
+d_3^{(3)} \,\bigg[\frac{1}{\nu-1}-\frac{1}{\nu +1}\bigg]+d_4^{(3)}.
\ee
Its large $\nu$ expansion is
\be
\la{4.22}
\wt F_{3}(\nu) = d_4^{(3)}+\frac{2 d_3^{(3)}}{\nu ^2}+\frac{6 d_2^{(3)}+2 d_3^{(3)}}{\nu ^4}
+\frac{10 d_1^{(3)}+20 d_2^{(3)}+2
   d_3^{(3)}}{\nu ^6}+\dots.
\ee 
From the quasi-modular expansion, we have 
\begin{align}
\la{4.23}
f_{0}(q) &= 4\,\log\prod_{n=1}^{\infty}(1-q^{2n}) = -4 \,q^2-6 \,q^4-\frac{16 \,q^6}{3}+\dots, \notag \\
\lambda_{2} &= \frac{1}{9} \left(\E_2^2-\E_4\right) = -32 \,q^2-192 \,q^4-384 \,q^6+\dots, \\
\lambda_{3} &= \frac{2}{135} \left(5 \E_2^3-6 \E_4 \E_2+\E_6\right) = -32 \,q^2+192 \,q^4+3456 \,q^6+\dots, \notag \\
\lambda_{4} &= \frac{1}{567} \left(35 \E_2^4-49 \E_4 \E_2^2+12 \E_6 \E_2+2 \E_4^2\right) = 
-32 \,q^2+1088 \,q^4+7296 \,q^6+\dots.\notag
\end{align}
Hence, we must have 
\be
\la{4.24}
\wt F_{3}(\nu) = -\frac{16}{3}+\frac{32}{\nu^{2}}-\frac{288}{\nu^{4}}-\frac{608}{\nu^{6}}+\dots.
\ee
Comparing (\ref{4.24}) and (\ref{4.22}), we obtain
\be
d_{1}^{(3)}=-\frac{128}{3}, \quad d_{2}^{(3)}=-\frac{160}{3}, \quad d_{3}^{(3)}=16,
\quad d_{4}^{(3)}=-\frac{16}{3},
\ee
in agreement with $Q_{3}$ in (\ref{2.9}).

\section{The fate of $\Omega$-deformation singularities at all instantons: finite-gap}
\la{sec:poles-1}

In this Section, we shall discuss the $\Omega$-deformation singularity at $\nu=1$ 
at all instantons in the 1- and 2-gap cases, {\em i.e.} at $\mu=2$ and $6$ respectively.

\subsection{Full analysis of the 1-gap problem}

By applying the methods discussed in the previous Section, we have computed $\wt F_{k}(\nu)$
for instanton number $k$ up to 24. We remark that this is a non-trivial computational problem with other approaches
or for general mass and/or beyond the NS limit. 
This effort allows to analyze possible regularities in the Nekrasov functions, or equivalently
in the sets $\mathbf{d}^{(k)} = \{d_{1}^{(k)}, \dots, d_{k+1}^{(k)}\}$. From our explicit data  we could
check the following remarkable relations for the pole coefficients in (\ref{2.11}). After introducing
\be
\la{5.1}
\mc D_{k} = \frac{(-1)^{k+1}\,2^{2k}\,(2k-2)!}{(k!)^{2}},
\ee
we obtain 
\begin{align}
\la{5.2}
d_{1}^{(k)} &= \mc D_{k}, \qquad
d_{2}^{(k)} = -\frac{k\,(2k-1)}{4\,(2k-3)}\,\mc D_{k}\,\\
d_{3}^{(k)} &= \frac{k\,(4k^{3}-18k^{2}+23k-3)}{32\,(2k-3)\,(2k-5)}\,\mc D_{k},\notag \\
d_{4}^{(k)} &= -\frac{k\,(8 k^5-96 k^4+440 k^3-891 k^2+659 k-30)}{384\,(2k-3)\,(2k-5)\,(2k-7)}\,\mc D_{k},
\notag \\
d_{5}^{(k)} &= \frac{k\,( 16 k^7-368 k^6+3556 k^5-18239 k^4+51445 k^3-73787 k^2+40527 k-630)}
{6144\,(2k-3)\,(2k-5)\,(2k-7)\,(2k-9)}\,\mc D_{k},\notag \\
d_{6}^{(k)} &= -\frac{k}{122800\,(2k-3)(2k-5)(2k-7)(2k-9)(2k-11)}\,\bigg(
32 k^9-1200 k^8+19800 k^7 \notag \\
&-186330 k^6+1082901 k^5-3927855 k^4+8529985 k^3-9884655 k^2 \notag \\
&+4503402 k-22680
\bigg)\, \mc D_{k},\notag
\end{align}
The general pattern for $p>1$ is 
\be
d^{(k)}_{p} = \frac{(-1)^{p+1}\,k\,G_{p}(k)}{4^{p-1}\,\Gamma(p)\,\prod_{\ell=1}^{p-1}(2k-2\ell-1)}\,\mc D_{k},
\ee
where $G_{p}(k)$ is a polynomial with degree $2p-3$ and integer coefficients. Also, it is not hard to 
compute additional terms beyond those in (\ref{5.2}). We emphasize that these expressions are rather peculiar. They provide the exact dependence of certain features of the exact Nekrasov prepotential $\wt F_{k}$ 
in terms of  formulas that are parametric in $k$, {\em i.e.} that hold for all $k$. This is 
quite an important fact because it opens the way to 
possible resummations over all instantons. This is precisely what we are going to do.

\subsubsection{Summing the singular poles over all instantons}

The total instanton partition function is
\be
\la{5.4}
\wt F(\nu, q) = \sum_{k=1}^{\infty}\wt F_{k}(\nu)\,q^{2k}.
\ee
As a first step, we analyze its singular part  in the  $\nu\to 1$ limit. From (\ref{2.11}), this is 
\be
\la{5.5}
\begin{split}
\wt F^{\rm \,sing}(\nu, q) &= \sum_{k=1}^{\infty}\sum_{p=1}^{k} \frac{d_{p}^{(k)}\,q^{2k}}{(\nu-1)^{2k-2p+1}} = 
\sum_{p=1}^{\infty}\sum_{k=p}^{\infty}\frac{d_{p}^{(k)}\,q^{2k}}{(\nu-1)^{2k-2p+1}} \\
& =\sum_{p=1}^{\infty}\,q^{2p-1}\,\sum_{k=p}^{\infty}\frac{d_{p}^{(k)}\,q^{2k-2p+1}}{(\nu-1)^{2k-2p+1}} 
%= \sum_{p=1}^{\infty}\,q^{2p-1}\,\sum_{k=0}^{\infty}\frac{d_{p}^{(k+p)}\,q^{2k+1}}{(\nu-1)^{2k+1}} 
= \sum_{k=1}^{\infty}\,q^{2k-1}\,\sum_{p=0}^{\infty} d_{k}^{(k+p)}\,\left(\frac{q}{\nu-1}\right)^{2p+1} \\
&= \sum_{k=1}^{\infty}\,g_{k}\left(\frac{q}{\nu-1}\right)\, q^{2k-1}, 
\end{split}
\ee
where we have introduced the functions
\be
\la{5.6}
g_{k}(z) =  \sum_{p=0}^{\infty} d_{k}^{(k+p)}\,z^{2p+1}.
\ee
Let us look at the term with $k=1$. This is not the pure 1-instanton term because the argument of the
functions $g_{k}$ have an argument that also depends on $q$. An explicit summation gives $g_{1}$
in terms of an hypergeometric function
\be
\la{5.7}
g_{1}(z) = 4\,z\,\,{}_{3}F_{2}(\tfrac{1}{2}, 1, 1; 2, 2; -16\,z^{2}).
\ee
We can now take the  $z\to +\infty$ limit and we find 
\be
g_{1}(z) = 4+\mc O(1/z),
\ee
where the omitted terms have also logarithmically enhanced term $\sim \log z/z^{n}$. A similar result is obtained
for the other functions $g_{k}(z)$. In particular, we find 
\be
\la{5.9}
\begin{split}
g_{2}(z) &= \frac{32 \sqrt{16 z^2+1} z^2-24 z^2-\sqrt{16 z^2+1}+1}{24 z^3} = \frac{16}{3}+\mc O(1/z), \\
g_{3}(z) &=  \frac{-480 z^4-80 z^2+1}{320 z^5}+\frac{6144 z^6+1152 z^4+72 z^2-1}{320 z^5
   \sqrt{16 z^2+1}} = \frac{24}{5}+\mc O(1/z) , \\
g_{4}(z) &= \frac{7168 z^6+1344 z^4-168 z^2+1}{2688 z^7}\\
&+\frac{786432 z^{10}-270336
   z^8-23040 z^6+2592 z^4+144 z^2-1}{2688 z^7 \left(16 z^2+1\right)^{3/2}} = \frac{32}{7}+\mc O(1/z), \\
g_{5}(z) &= \frac{-87552 z^8+27648 z^6+8640 z^4-288 z^2+1}{18432 z^9}\\
&+\frac{1}{18432 z^9 \left(16 z^2+1\right)^{5/2}}\,\bigg(109051904
   z^{14}+38273024 z^{12}-21135360 z^{10}\\
   &-4794368 z^8-236288 z^6+2400 z^4+248
   z^2-1\bigg) = \frac{52}{9}+\mc O(1/z).
\end{split}
\ee
Collecting these results and adding similar computations for $g_{6}(z)$ and $g_{7}(z)$ we get
\be
\la{5.10}
\wt F^{\rm \, sing}(1, q) = 4\, q+\frac{16}{3}\,q^{3}+\frac{24}{5}\,q^{5}+\frac{32}{7}\,q^{7}
+ \frac{52}{9}\,q^{9}+\frac{48}{11}\,q^{11}+\frac{56}{13}\,q^{13}+\dots.
\ee
Remarkably, this expression takes the form of a {\em finite} series in $q$. \footnote{
The fact that (\ref{5.10}) is organized
in {\em odd} powers of $q$ is not in contradiction with (\ref{5.4}). The Taylor expansion of
the functions $g_{k}(z)$ around $z=0$ involves odd powers of $z$, 
see (\ref{5.6}), but here we are taking the large $z\to +\infty$ expansion.}
Besides, one can check that the expansion 
(\ref{5.10}) agrees with 
\be
\la{5.11}
\boxed{
\wt F^{\rm \, sing}(1, q) = 2\,\log\frac{\prod_{n=1}^{\infty}(1-(-q)^{n})}{\prod_{n=1}^{\infty}(1-q^{n})}
= -\frac{1}{4}\,\log(1-m), \qquad m=m(q).}
\ee 
Actually, (\ref{5.11}) is not a guess, but has a natural origin  from the Lam\'e equation as we shall 
discuss in a moment. Here, we just add that using the known relation between $m$ and $q$, 
\be
\frac{dq}{dm} = \frac{\pi^{2}\, q}{4\, m\,(1-m)\, \mathbb{K}^{2}},
\ee
it is possible to rewrite the $q\,\partial_{q}$ derivative of the singular part  (\ref{5.11}) 
in the more suggestive form 
\be
\la{5.13}
q\, \frac{d}{dq}\, \wt F^{\rm \, sing}(1,q) = \frac{m}{4}\, \bigg(\frac{2\,\mathbb K}{\pi}\bigg)^{2}.
\ee

\subsubsection{Summing the regular part over all instantons}

A completely similar computation can be repeated for the  part of (\ref{5.4}) that is 
regular when $\nu\to 1$.
This is closely related to the singular part due to the exact partial fraction decomposition  (\ref{2.11}).
Apart from the $d^{(k)}_{k+1}$ constant term, the contributions proportional to powers of $1/(\nu+1)$
give, see (\ref{5.5}) and  (\ref{4.18})
\be
\la{5.14}
\begin{split}
\wt F^{\rm \, reg}(1,q) &= \wt F^{\rm \, reg (I)}(q) +\wt F^{\rm \, reg (II)}(q), \\
\wt F^{\rm \, reg (I)}(q) &=- \sum_{k=1}^{\infty} g_{k}\left(\frac{q}{2}\right)\, q^{2k-1}, \qquad
\wt F^{\rm \, reg (II)}(q) = 4\,\log\prod_{n=1}^{\infty}(1-q^{2n}).
\end{split}
\ee
The explicit computation of the functions $g_{k}(z)$ that we did in (\ref{5.7}) and (\ref{5.9}) allows
 to evaluate 
the non-trivial part of  (\ref{5.14})  as follows
\be
\la{5.15}
\begin{split}
\wt F^{\rm \, reg (I)}(q)&= 
-q\,\bigg(2\, q-q^3+\frac{4\, q^5}{3}-\frac{5\, q^7}{2}+\frac{28\, q^9}{5}+\dots\bigg)\\
&-q^{3}\,\bigg(6 \,q-\frac{20 \,q^3}{3}+14 \,q^5-36 \,q^7+\dots\bigg)
-q^{5}\,\bigg(8 \,q-19 \,q^3+\frac{324 \,q^5}{5}+\dots\bigg)\\
&-q^{7}\,\bigg(14 \,q-44 \,q^3+\dots\bigg)
-q^{9}\,\bigg(12\,\,q+\dots\bigg)+\mc O(q^{12})\\
&= -2 \,q^2-5 \,q^4-\frac{8 \,q^6}{3}-\frac{13 \,q^8}{2}-\frac{12
   \,q^{10}}{5}+\mc O(q^{12}).
\end{split}
\ee
Adding the second term in (\ref{5.14}), the full regular part of $\wt F(\nu)$ at $\nu=1$ is thus
\be
\la{5.16}
\wt F^{\rm \, reg}(1,q)  = -6 \,q^2-11 \,q^4-8 \,q^6-\frac{27 \,q^8}{2}-\frac{36 \,q^{10}}{5}
-\frac{44 \,q^{12}}{3}-\frac{48 \,q^{14}}{7}-\frac{59 \,q^{16}}{4}-\frac{26
   \,q^{18}}{3}+\dots,
\ee
where we have included  a few additional terms beyond those written in (\ref{5.15}).

\subsubsection{All instanton partition function at $\nu=1$ from the Lam\'e equation}

At this point we can understand what is happening. Summing the singular poles and the 
regular part of $\wt F(1)$ we obtain from (\ref{5.10}) and (\ref{5.16}) the total result
\begin{align}
\la{5.17}
\wt F(1,q) &= 4 q-6 q^2+\frac{16 q^3}{3}-11 q^4+\frac{24 q^5}{5}-8 q^6+\frac{32
   q^7}{7}-\frac{27 q^8}{2}+\frac{52 q^9}{9}-\frac{36 q^{10}}{5}\\
   &+\frac{48
   q^{11}}{11}-\frac{44 q^{12}}{3}+\frac{56 q^{13}}{13}-\frac{48
   q^{14}}{7}+\frac{32 q^{15}}{5}-\frac{59 q^{16}}{4}+\frac{72
   q^{17}}{17}-\frac{26 q^{18}}{3}+\dots.\notag
\end{align}
Comparing with (\ref{3.14}), this means
\begin{align}
\la{5.18}
\Lambda&(\nu=1, q) = -2\,q\,\frac{d}{dq}\,\bigg[\wt F(1,q)-8\,\sum_{n=1}^{\infty}\log(1-q^{2n})\bigg] = 
-8 \,q-8 \,q^2-32 \,q^3-8 \,q^4-48 \,q^5\notag \\
&-32 \,q^6-64 \,q^7-8 \,q^8-104 \,q^9-48 \,q^{10}-96
   \,q^{11}-32 \,q^{12}-112 \,q^{13}\\
   &-64 \,q^{14}-192 \,q^{15}-8 \,q^{16}-144 \,q^{17}-104
   \,q^{18}+\mc O(q^{19}).\notag
\end{align}
However, the exact spectrum of the Lam\'e equation predicts that $\Lambda(\nu=1,q)$ is nothing but
$\lambda+\tfrac{1}{3}$ where $\lambda$ is the eigenvalue of the scattering band edge 
associated with the potential (\ref{3.5}). For the potential $V(x)=2\,m\,\text{sn}^{2}(x,m)$ 
this is simply $m+1$.
Taking into account the sign of the eigenvalue as defined in (\ref{4.7}), as well as the scaling factor to 
map the equation on $x\in [0,2\pi]$, we have the exact result
\be
\la{5.19}
\Lambda(\nu=1,q) = \frac{1}{3}-\bigg[2\,\bigg(-\frac{m+1}{3}\bigg)+m+1\bigg]\,\bigg(
\frac{2\,\mathbb K}{\pi}\bigg)^{2} = \frac{1}{3}-\frac{m+1}{3}\,\bigg(\frac{2\,\mathbb K}{\pi}\bigg)^{2}
\ee
The expansion of this quantity gives indeed 
\be
\begin{split}
& \frac{1}{3}-\frac{m+1}{3}\,\bigg(\frac{2\,\mathbb K}{\pi}\bigg)^{2} = -\frac{m}{2}-\frac{9 m^2}{32}-\frac{13 m^3}{64}-\frac{1321
   m^4}{8192}-\frac{2207 m^5}{16384}-\frac{30461
   m^6}{262144}+\dots,
   \end{split}
\ee
and replacing $m$ by $q$ using (\ref{3.4}) we recover precisely (\ref{5.18}). This agreement is a check 
of our calculations and, in particular, of the closed expressions in (\ref{5.2}).

\subsection{Singular poles for the 2-gap problem}

A similar analysis can be worked out at the 2-gap point 
$\mu=2\times3=6$. In this case, the partial fraction expansion (\ref{2.11}) turns out to be just slightly more complicated and reads
\be
\la{5.21}
\begin{split}
\wt F_{k}(\nu) &= d_{k+1}^{(k)}+\sum_{p=1}^{k}d_{p}^{(k)}\,\bigg(\frac{1}{(\nu-1)^{2k-2p+1}}-\frac{1}{(\nu+1)^{2k-2p+1}}\bigg)\\
&+\sum_{p=1}^{ \lfloor \frac{k}{2} \rfloor }\,c^{(k)}_{p}\,\bigg(
\frac{1}{(\nu-2)^{2 \lfloor \frac{k}{2} \rfloor-2p+1}}-\frac{1}{(\nu+2)^{2 \lfloor \frac{k}{2} \rfloor-2p+1}}
\bigg).
\end{split}
\ee
This reflects the fact that there is a new set of Nekrasov poles located  at $\nu=\pm 2$. Other poles are 
always absent reflecting the special finite-gap choice.
We can repeat the analysis we did for the 1-gap case. In particular, it is interesting to study the behaviour of the 
singularity at the previous point $\nu=1$ that is still singular. Again, it is possible to derive closed expressions for the coefficients
$d_{p}^{(k)}$. The explicit results for $\mu=6$ (replacing the $\mu=2$ results in (\ref{5.2}))
are now similar, but slightly different
\begin{align}
\la{5.22}
d_{1}^{(k)} &= \mc D_{k}, \qquad
d_{2}^{(k)} = \frac{k\,(14k-23)}{36\,(2k-3)}\,\mc D_{k}\,\\
d_{3}^{(k)} &= \frac{k\,(196k^{3}+430k^{2}-3737k+3597)}{2592\,(2k-3)\,(2k-5)}\,\mc D_{k},\notag \\
d_{4}^{(k)} &= \frac{k\,(2744 k^5+31584 k^4-171640 k^3-287133 k^2+1734725 k-1375890)}{279936\,(2k-3)\,(2k-5)\,(2k-7)}\,\mc D_{k},
\notag
\end{align}
and so on, where now
\be
\la{5.23}
\mu=6: \qquad 
\mc D_{k} = -\frac{(-9)^{k}\,4^{2k-1}\,\Gamma(k-\tfrac{1}{2})}{\sqrt{\pi}\, k^{2}\, \Gamma(k)}.
\ee
The singular part of the full instanton partition function turns out to be compatible with the Ansatz
\be
\la{5.24}
\mu=6: \qquad \wt F^{\rm \, sing}(1) = -\frac{3}{4}\,\log(1-m).
\ee 
Thus, apart from a $\mu$-dependent normalization, this singular part
has a the same dependence on $m$ as in the 1-gap case.  In the next section, 
we shall explain that this is an accidental fact and provide the general case.
We remark that the finiteness of $\wt F^{\rm \, sing}(1)$  is ultimately due to the finiteness of the 
Lam\'e eigenvalues in terms of the quasi-momentum, on the edges of the spectral gaps.
This is essentially the function $\Lambda(\nu, q)$. However, 
$\wt F^{\rm\, sing}(1)$ is only a part of it, because $\Lambda$ includes the contributions from the 
other poles as well as the regular parts. 

\section{Beyond finite-gap, generic $\mu$}
\la{sec:non-gap}

The pattern in (\ref{5.1}, \ref{5.2}) and (\ref{5.22}, \ref{5.23}) suggests a universal structure in $k$
with the fine details being dependent on the parameter $\mu$. Guided by this results,
we analyzed the $\nu=1$ pole at generic $\mu$. Writing again 
\be
\la{6.1}
\wt F_{k}(\nu) = \sum_{p=1}^{k} \frac{d_{p}^{k}}{(\nu-1)^{2k-2p+1}}+\dots, 
\ee
After some work, we confirm that the pattern is indeed rather simple. Indeed, for generic $\mu$ we obtain 
\begin{align}
\la{6.2}
& \wt F_{k}^{\rm \, sing}(\nu)  = \frac{(-4)^{k-1}\,\mu^{2k}\,\Gamma(k-\tfrac{1}{2})}{\sqrt\pi\,k^{2}\,\Gamma(k)}
\,\bigg\{
\frac{1}{(\nu-1)^{2k-1}} \notag \\
&+\frac{k\,[k\,(3\mu^{2}-8\mu-4)-4(\mu^{2}-2\mu-1)]}{4\,\mu^{2}\,(2k-3)}\,\frac{1}{(\nu-1)^{2k-3}}\\
&+\frac{k}{576\,\mu^{4}\,(2k-3)(2k-5)}\,[
18 k^3 \,(3 \mu ^2-8 \mu -4)^2+k^2\, (-289 \mu ^4+2992 \mu ^3-3208
   \mu ^2\notag\\
   &-5184 \mu -1296)
   +k\, (-629 \mu ^4-2128 \mu ^3+4216 \mu
   ^2+7488 \mu +1872)\notag\\
   &+864 (\mu ^4-2 \mu ^2-4 \mu -1)
]\,
\frac{1}{(\nu-1)^{2k-5}}+\dots
\bigg\},\notag 
\end{align}
and so on. 
Of course, this expression reproduces the special 1- and 2-gap cases. However, (\ref{6.2}) shows that 
the all-instanton structure at generic $\mu$ is essentially the same.
Evaluating $\wt F^{\rm\, sing}(1,q)$, we get the remarkably simple expansion 
\footnote{We stress again that $\wt F^{\rm\, sing}(1,q)$ is obtained from 
 $\wt F^{\rm \, sing}(\nu, q)=\sum_{k}\wt F_{k}^{\rm \, sing}(\nu)\,q^{2k}$, doing the sum over $k$, i.e. over all instanton, and taking the limit $\nu\to 1^{+}$.}
\begin{align}
\la{6.3}
&\wt F^{\rm \, sing}(1,q) = 2\,\mu\,q-\frac{\mu}{6}\,(\mu^{2}-8\,\mu-4)\,q^{3}\notag\\
&\  +\frac{\mu}{360}\,(11\,\mu^{4}-80\,\mu^{3}+8\,\mu^{2}+576\,\mu+144)\,q^{5}\\
&\  +\frac{\mu}{8064}\,(55 \,\mu ^6-1232 \,\mu ^5+8568 \,\mu ^4-21824 \,\mu ^3+15472 \,\mu ^2+13824 \,\mu +2304)\,q^{7}+\dots.\notag
\end{align}
In particular, at the $n$-gap points, we find \footnote{As we remarked, the 2-gap result is
3 times the 1-gap one. This accidental simplification will find its explanation in the relation (\ref{6.7}).}
\be
\la{6.4}
\begin{split}
\text{1-gap}\qquad & 
\wt F^{\rm \, sing}(1,q) = 4\, q+\frac{16}{3}\,q^{3}+\frac{24}{5}\,q^{5}+\frac{32}{7}\,q^{7}+\dots, \\
\text{2-gap}\qquad & 
\wt F^{\rm \, sing}(1,q) = 12\, q+16\, q^3+\frac{72}{5}\,q^{5}+\frac{96}{7}\,q^{7}+\dots, \\
\text{3-gap}\qquad & 
\wt F^{\rm \, sing}(1,q) = 24\, q-88\, q^3+\frac{16344}{5}\, q^5+\frac{192}{7}\,q^{7}+\dots, \\
\text{4-gap}\qquad & 
\wt F^{\rm \, sing}(1,q) = 40\, q-\frac{2360}{3}\, q^3+63048\, q^5
+\frac{13547840}{7}\,q^{7}+\dots.
\end{split}
\ee 
Actually, we now show that the expansion (\ref{6.3}) has a  
very simple interpretation from the point of view
of the spectrum of the Lam\'e equation. This is most clearly explained resorting again to the $n$-gap
cases. The bands for the potential $V(x) = n\,(n+1)\,m\,\text{sn}^{2}(x,m)$ are for $n=1,2,3$
\be
\begin{split}
\text{1-gap}\qquad & [E^{(1)}_{0}, E^{(1)}_{1}]\cup[E^{(1)}_{2},+\infty), \\
\text{2-gap}\qquad & [E^{(2)}_{0}, E^{(2)}_{1}]\cup[E^{(2)}_{2},E^{(2)}_{3}]
\cup[E^{(2)}_{4},+\infty), \\
\text{3-gap}\qquad & [E^{(3)}_{0}, E^{(3)}_{1}]\cup[E^{(3)}_{2},E^{(3)}_{3}]
\cup[E^{(3)}_{4},E^{(3)}_{5}]\cup[E^{(3)}_{6},+\infty).
\end{split}
\ee
The relevant edges for our discussion are the upper edge of the first bound band and the lower edge of the 
second bound (or scattering in 1-gap case) band. This is because we are looking at the singularities at 
$\nu=1$. The difference $E^{(n)}_{2}-E^{(n)}_{1}$ is the {\em width of the first gap}. For $n=1,2,3$, it is known that 
\be
\la{6.6}
\begin{split}
& E_{2}^{(1)}-E_{1}^{(1)} = m, \qquad E_{2}^{(2)}-E_{1}^{(2)} = 3\,m, \\
& E_{2}^{(3)}-E_{1}^{(3)} = 2 \sqrt{m^2-m+4}-2 \sqrt{4 m^2-7 m+4}+3 m.
\end{split}
\ee
Comparing the first line of (\ref{6.6}) with (\ref{5.13}) and (\ref{5.24}), it is natural to conjecture that 
\be
\la{6.7}
\boxed{
q\,\frac{d}{dq}\,\wt F^{\rm \, sing}(1,q) = \frac{E^{(n)}_{2}-E^{(n)}_{1}}{4} \,\left(
\frac{2\,\mathbb K}{\pi}\right)^{2}.}
\ee
%
%The explicit values are known to be 
%\be
%\begin{split}
%&E^{(1)}_{1} = 1, \qquad E^{(1)}_{2} = 1+m, \\
%&E^{(2)}_{1} = 2 \sqrt{m^2-m+1}-m-1, \qquad E^{(2)}_{2} = 2 \sqrt{m^2-m+1}+2 m-1, \\
%&E^{(3)}_{1} = -2 \sqrt{4 m^2-7 m+4}+2 \sqrt{4 m^2-m+1}+3, \\
%& E^{(3)}_{2} = -2 \sqrt{m^2-m+4}+2 \sqrt{4
%   m^2-m+1}-3 m+3.
%\end{split}
%\ee
%In particular,
%\be
%\frac{E^{(1)}_{2}-E^{(1)}_{1}}{4} = \frac{m}{4}, \qquad
%\frac{E^{(2)}_{2}-E^{(2)}_{1}}{4} = \frac{3\,m}{4}, 
%\ee
%and
%\be
%\frac{E^{(3)}_{2}-E^{(3)}_{1}}{4} = 
%\frac{1}{4} \left(2 \sqrt{m^2-m+4}-2 \sqrt{4 m^2-7 m+4}+3 m\right).
%\ee
%The general relation appears to be 
%\be\boxed{
%q\,\frac{d}{dq}\,\wt F^{\rm \, sing}(1,q) = \frac{E^{(n)}_{2}-E^{(n)}_{1}}{4} \,\left(
%\frac{2\,\mathbb K}{\pi}\right)^{2}.}
%\ee
Indeed and rather non-trivially, this also agrees with the 3-gap case.
In particular, it implies the following all-order expansion replacing the third row of (\ref{6.4})
\be
\begin{split}
\text{3-gap}\qquad & 
\wt F^{\rm \, sing}(1,q)  = 
24\, q-88\, q^3+\frac{16344}{5}\,q^{5}+\frac{192}{7}\,q^{7}-\frac{27856816}{3}\,q^{9}\\
   &+\frac{7624002168}{11}\,q^{11}-\frac{122964074664}{13}\,q^{13}
   -\frac{16297316089728}{5}\,q^{15}\\
   &+\frac{6101221040271792}{17}\,q^{17}
   -\frac{202648353741486600}{19}\,q^{19}+\dots.
\end{split}
\ee
Actually, a similar discussion can be done for generic $\mu$. To this aim, we need the 
expansion of the Lam\'e eigenvalues $E_{1}, E_{2}$ for generic $\mu$. This can be achieved 
efficiently by the methods presented in  \cite{volkmer2004four}. We did this exercise at high order in 
the modular parameter $m$. The first terms of the expansions are 
\begin{align}
\la{6.9}
E_{1} &= 1+\frac{1}{4} (\mu -2) m-\frac{1}{128} ((\mu -6) (\mu -2)) m^2+\frac{(\mu -16)
   (\mu -6) (\mu -2) m^3}{4096}\notag\\
   &-\frac{\left((\mu -6) (\mu -2) \left(\mu ^2-104
   \mu +972\right)\right) m^4}{393216}\notag \\
   &-\frac{\left((\mu -6) (\mu -2) \left(11
   \mu ^3+152 \mu ^2-9420 \mu +66240\right)\right) m^5}{37748736}\notag \\
   &+\frac{(\mu
   -6) (\mu -2) \left(49 \mu ^4-1608 \mu ^3-11668 \mu ^2+552960 \mu
   -3223296\right) m^6}{2415919104}+\dots, \notag \\
E_{2} &= 1+\frac{1}{4} (3 \mu -2) m-\frac{1}{128} ((\mu -6) (\mu -2)) m^2-\frac{((\mu
   -6) (\mu -2) (\mu +16)) m^3}{4096}\\
   &-\frac{\left((\mu -6) (\mu -2) \left(\mu
   ^2+88 \mu +972\right)\right) m^4}{393216}\notag \\
   &+\frac{(\mu -6) (\mu -2) \left(11
   \mu ^3-136 \mu ^2-7116 \mu -66240\right) m^5}{37748736}\notag \\
 &  +\frac{(\mu -6) (\mu
   -2) \left(49 \mu ^4+1208 \mu ^3-9620 \mu ^2-384000 \mu -3223296\right)
   m^6}{2415919104}+\dots.\notag
\end{align}
Their difference is somewhat simpler
\begin{align}
\la{6.10}
&E_{2}-E_{1} = \frac{\mu  m}{2}-\frac{(\mu -6) (\mu -2) \mu \, m^3}{2048}-\frac{(\mu -6)
   (\mu -2) \mu \, m^4}{2048}\\
   &+\frac{(\mu -6) (\mu -2) \mu  \left(11 \mu ^2+8
   \mu -8268\right) m^5}{18874368}+\frac{(\mu -6) (\mu -2) \mu  \left(11 \mu
   ^2+8 \mu -3660\right) m^6}{9437184}+\dots. \notag
 \end{align}
 Using (\ref{6.10}) and  (\ref{6.3}), we match perfectly (\ref{6.7}) at the considered order.
 This provides a non-trivial check that the  relation (\ref{6.7}) holds for any $\mu$.
 
 Actually, it seems quite natural to generalize (\ref{6.7}) to the following more general form 
\be
\la{6.11}
\boxed{
q\,\frac{d}{dq}\,\wt F^{\rm \, sing}(N,q) = \frac{W_{N}}{4} \,\left(
\frac{2\,\mathbb K}{\pi}\right)^{2},}
\ee
where $N=1, 2, \dots$, and $W_{N}$ is the width of the N-th gap of the Lam\'e spectrum.
A simple check of (\ref{6.11}) in the 2-gap case is indeed presented in App.~(\ref{app:new}). 
 
\section{Conclusions}

To summarize, we have shown that the $k$-instanton prepotential of $\mc N=2^{*}$ $SU(2)$
gauge theory has a nice structure in the $n$-gap Nekrasov-Shatashvili limit defined by 
\be
\eps_{2}=0, \qquad m = \bigg(n+\frac{1}{2}\bigg)\,\eps_{1}, \qquad n\in\mathbb N.
\ee
Up to trivial factors, it is a rational function of the ratio $\nu=\frac{2a}{\eps_{1}}$ where $a$ is the 
scalar field expectation value. This Nekrasov function $\wt F_{k}(\nu)$ has poles at 
$|\nu|=1, 2, \dots, n$ only. We studied the structure of this poles (and of the regular part) by 
exploiting the Bethe/gauge map finding simple closed expressions parametric in $k$.

The Bethe/gauge correspondence predicts that the Nekrasov functions may be obtained from the 
eigenvalues $\lambda$
of the associated spectral problem for the  Lam\'e equation. What is needed is simply
 the expression of
$\lambda$ as a function of the Floquet exponent to be identified with the above ratio $\nu$.
The exact expression $\lambda(\nu)$ is not singular at the Nekrasov poles $|\nu|=1,2,\dots, n$.
The poles are an artifact of the modular expansion in powers of the instanton counting 
coupling $q^{2}$. We explicitly clarified these relations. In particular, in the 1-gap case, 
we summed over all 
instantons and showed explicitly the cancellation of the Nekrasov poles. Adding the finite part
of the Nekrasov functions, we also checked 
agreement with 
the Lam\'e eigenvalue at the scattering band edge.

Focusing on the singular part only, we remark that the all-instanton finite resummation of the Nekrasov poles is a special contribution 
to the total Nekrasov function, not immediately related to quanitites appearing in the 
Lam\'e problem. As we explained in the 
Introduction, it is interesting
because it is expected to display some simple  behaviour. This is due to the fact that, 
at any instanton number, the poles are  associated with the structure of supersymmetric states in a 
quantum mechanical model that arises in the 5d $\Omega$-deformation construction.  
Actually, we showed that the resummed poles give a finite contribution at generic $m/\eps_{1}$ 
-- not necessarily finite-gap --  
and provided its clean Lam\'e equation interpretation in terms of the widths of the spectral gaps.

\section*{Acknowledgments}
We  thank  Y. Tachikawa, D. Krefl, 
%S. Hellerman, 
D. Orlando,  
%S. Reffert, 
and 
A. Zein Assi for important comments. We also thank D. Fioravanti and G. Macorini
for clarifying discussions.

\appendix

\section{Direct perturbative expansion of $\lambda(\nu)$ from the Lam\'e equation}
\la{A}

The expansion (\ref{3.11}) may be also obtained by a direct perturbative approach \cite{muller2006introduction}.
Just to show how it works, one simply writes the Lam\'e equation with rescaled variable $x\in[0,2\pi]$ 
and writes 
\be
\la{A.1}
\begin{split}
& \psi''(x)-2\,\bigg[-\frac{m+1}{3}+m\,\text{sn}^{2}\bigg(\frac{2\mathbb K}{\pi}\, x, m\bigg)\bigg]\,\psi(x) = 
\lambda\,\psi(x), \\
& \lambda = \frac{2}{3}-\nu^{2}+\lambda_{1}\,m+\lambda_{2}\,m^{2}+\lambda_{3}\,m^{3}+\dots.
\end{split}
\ee
The $\mc O(m^{3})$ wave function with definite quasi-momentum may be found by some calculation and it is 
\begin{align}
\la{A.2}
\psi(x) &= e^{i\,\nu\,x}\bigg[
1-\frac{m}{2}+\frac{i\,\nu\,m}{4\,(\nu^{2}-1)}\,\bigg(1+\frac{21\nu^{4}-46\nu^{2}+9}{256\,(\nu^{2}-1)^{2}}
\,m^{2}+
\bigg)\, \sin(2x)\\
&
-\frac{m}{4\,(\nu^{2}-1)}\,\bigg(1+\frac{23\nu^{4}-58\nu^{2}+19}{256(\nu^{2}-1)^{2}}\,m^{2}\bigg)\,
\cos(2x)\notag \\
&+\frac{i\,\nu\,m^{2}}{64(\nu^{2}-1)}\,\bigg(1+\frac{m}{2}\bigg)\,\sin(4x)\notag \\
&-\frac{m^{2}}{64\,(\nu^{2}-1)}\,\bigg(1+\frac{m}{2}\bigg)\,\cos(4x)
+\frac{i\,\nu\,m^{3}}{1024\,(\nu^{2}-1)}\,\sin(6x) \notag \\
&-\frac{m^{3}}{1024\,(\nu^{2}-1)}\,\cos(6x)+\mc O(m^{4})
\bigg].\notag
\end{align}
Replacing (\ref{A.2}) in (\ref{A.1}) determines $\lambda_{1},\lambda_{2},\lambda_{3}$ in agreement with 
(\ref{3.11}). 
The structure is clear, but going to very high orders in $m$ is rather cumbersome, whereas the approach described in the 
main text and based on the representation (\ref{3.9}) is fully straightforward and bypass the construction of $\psi(x)$.

\section{Resummation of singular 2-gap poles at $\nu=2$}
\la{app:new}

At $\mu=2\times 3=6$, the general structure of Nekrasov functions is (\ref{5.21}).
%(notation of \cite{Beccaria:2016wop})
%\be
%\la{6.2.1}
%\begin{split}
%\wt F_{k}(\nu) &= d_{k+1}^{(k)}+\sum_{p=1}^{k}d_{p}^{(k)}\,\bigg(\frac{1}{(\nu-1)^{2k-2p+1}}-\frac{1}{(\nu+1)^{2k-2p+1}}\bigg)\\
%&+\sum_{p=1}^{ \lfloor \frac{k}{2} \rfloor }\,c^{(k)}_{p}\,\bigg(
%\frac{1}{(\nu-2)^{2 \lfloor \frac{k}{2} \rfloor-2p+1}}-\frac{1}{(\nu+2)^{2 \lfloor \frac{k}{2} \rfloor-2p+1}}
%\bigg).
%\end{split}
%\ee
An analysis of the terms proportional to inverse powers of $\nu\pm 2$ gives
for $k\ge 1$
\be
\la{6.2.2}
\begin{split}
\text{even}\ k: & \quad c_{1}^{(k)} = \mc D^{(+)}_{k} \equiv 
\frac{(-1)^{\frac{k}{2}+1}\,24^{k}\,\Gamma(\frac{k-1}{2})}{\sqrt\pi\,k^{2}\,\Gamma(\frac{k}{2})}, \\
\text{odd}\ k: & \quad c_{1}^{(k)} = \mc D^{(-)}_{k} \equiv 
\frac{(-1)^{\frac{k-1}{2}}\,3^{k-1}\,2^{3k+1}\,\Gamma(\frac{k}{2}-1)}{\sqrt\pi\,\Gamma(\frac{k+1}{2})}.
\end{split}
\ee
At next-to-leading order, we obtain the following rational corrections for $k\ge 4$
\be
\la{6.2.3}
\begin{split}
\text{even}\ k: & \quad c_{2}^{(k)} = -\mc D^{(+)}_{k}\, 
\frac{k \left(256 \,k^2+473 \,k+119\right)}{288 \,(k-3)},\\
\text{odd}\ k: & \quad c_{2}^{(k)} = -\mc D^{(-)}_{k}\,
\frac{256 \,k^3+1163 \,k^2+941 \,k+520}{864 \,(k-4)}.
\end{split}
\ee
The pattern continues in the same way. For $k\ge 6$, 
\be
\la{6.2.4}
\begin{split}
&\text{even}\ k:  \\
&\quad c_{3}^{(k)} = \mc D^{(+)}_{k}\, 
\frac{k \left(65536 k^5+595456 k^4+1451987 k^3+1929413 k^2+1572789 k+367182\right)}{497664 (k-5)
   (k-3)} \\
&\text{odd}\ k:\\
& \quad c_{3}^{(k)} = \mc D^{(-)}_{k}\,
\frac{65536 k^6+948736 k^5+4141343 k^4+9278505 k^3+14843229 k^2+9404819 k+3250632}{2488320 (k-6)
   (k-4)}
\end{split}
\ee
The resummation of the poles is then
\be
\la{6.2.5}
\begin{split}
\mathop{\sum_{k=2}^{\infty}}_{\text{even}\ k} c_{1}^{(k)}\,\frac{q^{2k}}{(\nu-2)^{k-1}} &=
\frac{144\,q^{4}\,{}_{3}F_{2}(\frac{1}{2},1,1; 2,2; -\frac{576\,q^{4}}{(\nu-2)^{2}})}{\nu-2}
=24\,q^{2}+\mc O(\nu-2), \\
\mathop{\sum_{k=3}^{\infty}}_{\text{odd}\ k} c_{1}^{(k)}\,\frac{q^{2k}}{(\nu-2)^{k-2}} &=
-768\,q^{4}+\mc O(\nu-2), 
\end{split}
\ee
\be
\la{6.2.6}
\begin{split}
\mathop{\sum_{k=4}^{\infty}}_{\text{even}\ k} c_{2}^{(k)}\,\frac{q^{2k}}{(\nu-2)^{k-3}} &=
61472\,q^{6}+\mc O(\nu-2), \\
\mathop{\sum_{k=5}^{\infty}}_{\text{odd}\ k} c_{2}^{(k)}\,\frac{q^{2k}}{(\nu-2)^{k-4}} &=
-6703104\,q^{8}+\mc O(\nu-2), 
\end{split}
\ee
\be
\la{6.2.7}
\begin{split}
\mathop{\sum_{k=6}^{\infty}}_{\text{even}\ k} c_{3}^{(k)}\,\frac{q^{2k}}{(\nu-2)^{k-5}} &=
\frac{4346265744}{5}\,q^{10}+\mc O(\nu-2), \\
\mathop{\sum_{k=7}^{\infty}}_{\text{odd}\ k} c_{3}^{(k)}\,\frac{q^{2k}}{(\nu-2)^{k-6}} &=
-125613343744\,q^{12}+\mc O(\nu-2), 
\end{split}
\ee
and so on. The total singular part at $\nu=2$ is therefore
\be
\begin{split}
\wt F^{\rm sing}(\nu=2) &= 24\,q^{2}-768\,q^{4}+61472\,q^{6}-6703104\,q^{8}\\
&
+\frac{4346265744}{5}\,q^{10}-125613343744\,q^{12}+\dots.
\end{split}
\ee
Applying $q\partial_{q}$ and expressing the result in terms of the modular parameter $m$ we get
\be
\begin{split}
q\partial_{q}\,\wt F^{\rm sing}(\nu=2) &=
\frac{3 m^2}{16}+\frac{3 m^3}{16}+\frac{63 m^4}{512}+\frac{15 m^5}{256}+\frac{3585
   m^6}{131072}+\frac{3843 m^7}{131072}+\frac{183711 m^8}{4194304}\\
   &+\frac{52527
   m^9}{1048576}+\frac{360024561 m^{10}}{8589934592}+\frac{235184565
   m^{11}}{8589934592}\\
   &+\frac{5020080441 m^{12}}{274877906944}+\frac{2640895803
   m^{13}}{137438953472}+\mc O(m^{14}).
\end{split}
\ee
The width of the second gap of the Lam\'e equation at $\mu=6$ is 
\be
W_{2} = 2\,\sqrt{m^{2}-m+1}+m-2.
\ee
One checks that (\ref{6.11}) holds, {\em i.e.}
\be
q\partial_{q}\,\wt F^{\rm sing}(\nu=2) = \frac{1}{4}\,W_{2}\,\left(\frac{2\,\mathbb K}{\pi}\right)^{2}.
\ee

\section{Modular functions}
\la{app:eisen}

The Eisenstein series appearing in the main text may be defined by the expansions
\be
\begin{split}
\E_{2}(q) &= 1-24\,\sum_{n=1}^{\infty}\frac{n\,q^{n}}{1-q^{n}}, \ \ \ \
\E_{4}(q) = 1+240\,\sum_{n=1}^{\infty}\frac{n^{3}\,q^{n}}{1-q^{n}}, \ \ \ \
\E_{6}(q) = 1-504\,\sum_{n=1}^{\infty}\frac{n^{5}\,q^{n}}{1-q^{n}}.
\end{split}
\ee
The series $\E_{4}$ and $\E_{6}$ are true modular forms of 
weight 4 and 6. Under $SL(2, \mathbb{Z})$ modular transformation
\be
\tau\to \tau' = \frac{a\,\tau+b}{c\,\tau+d}, \qquad a,b,c,d\in\mathbb Z, \ \ \text{with}\ a\,d-b\,c=1,
\ee
they transform as 
\be
\E_{4}(\tau') = (c\,\tau+d)^{4}\,\E_{4}(\tau), \qquad
\E_{6}(\tau') = (c\,\tau+d)^{6}\,\E_{4}(\tau).
\ee
The series $\E_{2}(q)$ is a quasi-modular form of weight 2, with the transformation property
\be
\E_{2}(\tau') = (c\,\tau+d)^{2}\, \E_{2}(\tau)+\frac{6}{i\,\pi}\,c\,(c\,\tau+d).
\ee
We also remind  the definition of the Dedekind $\eta$-function
\be
\eta(q) = q^{\frac{1}{12}}\,\prod_{n=1}^{\infty}(1-q^{2n}).
\ee
Its modular properties may be found, for instance,  in \cite{koblitz2012introduction}.

\bibliography{N2-Biblio}
\bibliographystyle{JHEP}

\end{document}